# How false vacuum synthesis of a universe sets initial conditions which permit the onset of variations of a nucleation rate per Hubble volume per Hubble time.

A. W. Beckwith
Department of Physics and Texas Center for Superconductivity and Advanced
Materials at the University of Houston
Houston, Texas 77204-5005, USA

**ABSTRACT**

Using Bogomil'nyi inequality and the vanishing of topological charge at the onset of nucleation of a new universe permits a simpler, more direct insight into how topological defects (kinks and anti kinks) contribute to initial conditions at the onset of inflationary cosmology . Currently, there are few bridges between initial conditions for cosmological inflation and the nucleation of a new universe. This presentation shows how this can be done while still employing Venezianos prescription for forming a link between quanta of length, the magnitude of a dilaton field $\phi$ and forces gravitational and gauge alike

**Correspondence:** A. W. Beckwith: **projectbeckwith2@yahoo.com**.

PAC numbers   02.30. Em, 03.75.Lm, 11.27.+d, 98.65.Dx , 98.80.-k

# I. INTRODUCTION

In this paper, I offer a different paradigm for how topological defects (kinks and anti kinks) contribute to the onset of initial conditions at the beginning of inflationary cosmology. Mark Trodden[1] and Trodden et al[2] view topological defects as similar to D branes of string theory[1]. While this soliton-antisoliton (S-S') construction permits extensions to various super symmetric theories, it obscures direct links to inflationary cosmological potentials such as Guth's harmonic potential.[3]

Instead, to make the linkage clearer, I use S-S' pairs as an initial starting point for times $t \leq t_P$, where $t_P$ is Plancks discretized smallest unit of time as a coarse graining of time-stepping in cosmological evolution. In this paper, we start with a driven Sine Gordon potential,[4] which has much of its potential contribution eliminated in the times immediately after $t_P$, due to the vanishing of topological charge Q.[5] Afterwards, a second potential regime initiates that is proportional to the square of a scalar field $\phi$ divided by a denominator of the form $1 + A \cdot \phi^3$ [6,7], which in turn blends into the square of a scalar field $\phi$ for a third potential regime corresponding to Guth's chaotic inflationary model.[3] This second potential regime is akin to what happens in typical inflationary models, with an initially large potential forming a potential barrier at the onset of time $t \approx t_P + \delta \cdot t$ — and then decreasing into the third potential regime proportional $\phi^2$ for times $t >> t_P$ [3] and can be observed in Dodelson's[8] Figure 6.6 (scalar field trapped in a false vacuum).

There are three premises. First, for times less than or equal to Planck time $t_P$, the potential system for analyzing the nucleation of a universe is a driven Sine Gordon system[9], with the driving force in magnitude far less than the overall classical Sine

Gordon potential. The second premise is that topological charges for a S-S' stem occur prior to Planck time $t_P$ for this potential system to cancel out, leaving a potential that is proportional to $\phi^2$ minus a contribution due to quantum fluctuations of a scalar field that are equal in magnitude to a classical system, with the remaining scalar potential field contributing to cosmic inflation in the history of the early universe. The third premise is that minimum value for a vacuum fluctuation of energy equivalent to $\Delta t \cdot \Delta E = \hbar$ will lead to the nucleation of a new universe, provided that we are setting our initial time $t_P \approx \Delta t$ as the smallest amount of time which can be ascertained in a quantum universe.

These three premises permit a simpler and more direct specification of set values for the scale factor $a$ at the onset of inflation, namely at the end of the nucleation of a template for an expansionary universe. More importantly, we are finding a way to give appropriate initial conditions for the nucleation rate per Hubble-volume per Hubble-time that is conveniently set proportional to unity. This is done in the context of a flat universe, as well as allowing us to specify, in a more direct manner, starting points for the relative radius of a universe at the onset of cosmic inflation, as well as the behavior of the Hubble parameter in the onset of inflation. I also propose that, in contrast to contemporary models, this approach provides information about how quantum vacuum energy fluctuations affect the evolvement of a soliton kink–antikink contribution to cosmological inflation.

We need to keep in mind that if a phase transition occurs right after our nucleation of an initial state, the time of nucleation is actually less than (or equal to) Plancks minimum time interval $t_P$, with the length specified by reconciling the fate of the false vacuum potential used in nucleation with a Bogomol'nyi inequality specifying the vanishing of topological charge. I use S-S' pairs to represent an initial scalar field, which after time $t_P \approx \Delta t$ will descend into the typical chaotic inflationary potential used for inflationary cosmology.

## II. CHAOTIC INFLATIONARY SCENARIOS AND THEIR TIE IN WITH OUR PROBLEM OF FALSE VACUUM NUCLEATION

In Guth's[3] recent discussions of the basic workings of inflationary models, the simplest model is the chaotic inflationary model.[3] Via use of a massive scalar field construction, Guth's elegant treatment suggests how we could have an inflation field $\phi$ set at a high value $\phi \equiv \widetilde{\phi}_0$ and which, then, would have an inequality of[3]

$$\widetilde{\phi}_0 > \sqrt{\frac{60}{2 \cdot \pi}} M_P \approx 3.1 M_P. \tag{1}$$

This pre supposes a harmonic style potential of the form[3]

$$V \equiv \frac{1}{2} \cdot m^2 \cdot \phi^2 \tag{2}$$

where we have classical and quantum fluctuations approximately giving the same value for a phase value of[3]

$$\phi^* \equiv \left(\frac{3}{16 \cdot \pi}\right)^{\frac{1}{4}} \cdot \frac{M_P^{3/2}}{m^{\frac{1}{2}}} \cdot M_P \rightarrow \left(\frac{3}{16 \cdot \pi}\right)^{\frac{1}{4}} \cdot \frac{M_P^{3/2}}{m^{\frac{1}{2}}} \tag{3}$$

where we have set $M_P$ as the typical Plancks mass that, in this paper, we normalized to being unity for the hybrid false vacuum-inflaton field cosmology example, as well as having set the general evolution of our scalar field as having the form of[3]

$$\phi \equiv \widetilde{\phi}_0 - \frac{m}{\sqrt{12 \cdot \pi \cdot G}} \cdot t \tag{4}$$

We assume that, after an interval of time greater than Plancks unit of time, a nucleation of an initial universe would be evolving toward chaotic inflationary cosmology parameter that is very similar to these given results.

## III. DESCRIPTION OF THE POTENTIAL USED FOR NUCLEATION AND ITS BLENDING INTO CHAOTIC INFLATIONARY COSMOLOGY.

I examine reasonable potentials that incorporate insights of the chaotic inflation model ( $\phi^2$ potential dependence ) with false vacuum nucleation. For this potential, I worked with[10,11]

$$V_1(\phi) = \frac{M_P^2}{2} \cdot (1 - \cos(\phi)) + \frac{m^2}{2} \cdot (\phi - \phi^*)^2 \qquad (5)$$

where $M_P > m$ as well as an overall potential of the form

$$V(\phi) \equiv \left[initial \ energy \ density\right] + V_1(\phi) \qquad (6)$$

where the *initial energy density* is a term from assuming a brane world type of potential usually written as[11]

$$V(\phi, \tilde{\psi}) = \frac{1}{4} \cdot (\tilde{\psi}^2 - M_P^2)^2 + \frac{1}{2} \cdot \lambda' \cdot \phi^2 \cdot \tilde{\psi}^2 + V_1(\phi) \qquad (7)$$

and where the radial component $\tilde{\psi}$ is nearly set equal to zero; and the scalar potential, in our case, is changed from a $\phi^2$ potential dependence to one where we incorporate a false vacuum nucleation procedure as given by $V_1(\phi)$. I look at setting values of $\phi \equiv \phi^{*\,1}$ due to the chaotic inflation model[3] and then consider a specific ratio of $M_P$ to mass m in order to work with this same value to which the inflaton field will lead. As seen in Fig. 1b, true and false vacuum minimum values[12] obtain when we are using the Bogomil'nyi inequality[5, 13] with

$$V_1(\phi_F) - V_1(\phi_T) \cong .373 \propto L^{-1} \tag{8}$$

namely

$$\frac{(\{\ \})}{2} \equiv \Delta E_{GAP} \equiv V_1(\phi_F) - V_1(\phi_T)$$

and

$$\{\ \} \equiv \{\ \}_A - \{\ \}_B \equiv 2 \cdot \Delta E_{GAP} \tag{9}$$

where I will, for this cosmological example, set :

$$\{\ \}_A \approx M_P^2 + 2 \cdot m^2$$
$$\{\ \}_B \approx \frac{2 \cdot M_P^2 \cdot \phi_T \cdot \phi_F}{3 \cdot !} \tag{10a,b}$$

I am assuming that the net topological charge will vanish and that, for a D+1 dimensional model with phenomenology equivalent to quasi one-dimensional behavior due to near instantaneous nucleation, I work with the situation as outlined in Fig. 1a where the quantity in brackets is set by Fig. 1b. The details of that pop up are such that I am assuming a toy model with a thin wall approximation to a topological S-S' pair equivalent to assuming that the false vacuum paradigm of Sidney Coleman[12] holds in the main part, (as well as Lee and Weinbergs topological solitons associated with a vacuum manifold SO(3) / U(1)[1,2 ,5, 14]).

The second part of this potentials behavior scales as

$$V_2(\phi) \approx \frac{(1/2) \cdot m^2 \phi^2}{(1 + A \cdot \phi^3)} \tag{11}$$

which corresponds to a decreasing scalar field $\phi$ right after times $t \geq t_P + \delta \cdot t$ and the beginning of the collapse of the total charge Q of this kink-antikink ensemble expanding into inflationary cosmological potential behavior. This then will lead to

$$V_3(\phi) \approx (1/2) \cdot m^2 \phi^2 \tag{12}{}^3$$

with the bridge between these three regimes of the form

$$\begin{array}{lll} V_1 & \rightarrow V_2 & \rightarrow V_3 \\ \phi(increase) \leq 2 \cdot \pi & \rightarrow \phi(decrease) \leq 2 \cdot \pi & \rightarrow \phi \approx \varepsilon^+ \\ t \leq t_P & \rightarrow t \geq t_P + \delta \cdot t & \rightarrow t \gg t_P \end{array} \tag{13}$$

## VI. PRESENTING A NEW WAY TO OBTAIN INITIAL EVOLUTION OF THE HUBBLE PARAMETER AND A RATE EQUATION

Garriga,[15] assuming a nearly flat De Sitter universe, also came up with an expression for the number density of particles per unit length (time independent )[15]

$$n \approx \frac{1}{2 \cdot \pi} \cdot \sqrt{M^2 + e \cdot \frac{E_0^2}{H^2}} \cdot \exp(-S_E) \tag{14}$$

where, for our purposes, we would set

$$M \leq M_P \rightarrow 1 \tag{15}$$

I prefer, instead, to use an estimation of a nucleation rate per Hubble-volume per Hubble-time[16]

$$\in(t) \equiv \lambda_0 / (H(t)^4) \approx 1 \tag{16}$$

to show the influence an evolving Hubble parameter would have, in early times, without the complexity of predicting the $S_E$ (a Euclidian action integral) that would be in our example a D+1 dimensional space knocked down to a quasi one-dimensional in character. I assume, also, rescaling of Planckian length to be unity where $\hbar \equiv c \equiv G \equiv 1$.

## V. PREDICTING HOW A SCALE FACTOR EVOLVES IN THE BEGINNING OF INFLATIONARY COSMOLOGY

Next, I look at how the scale factor, $a$, changes in time in a manner that will enable us to delineate Hubble parameter variations in the first few moments after creation. In doing this, we can note the typical value ( as given by Dodelson)[17]

$$a(t) \cong a_B \cdot \exp(H_B(t - t_B)) \qquad (17)$$

with $a_B$, $H_B$, and $t_B$ being scale factor, Hubble parameter, and time values at the end of an inflationary period of expansion. Needless to say, in doing this, we are not obtaining values of what the scale factor and Hubble parameter could be at the onset of inflation, which is a situation we wish to remedy. So we set

$$\tilde{a}_0 \equiv a_B \exp(H_B(t_P - t_B)) \qquad (18)$$

and approximate the evolution of phase after time $t_P$ via use of

$$\phi \equiv \tilde{\phi}_0 - \frac{m}{\sqrt{12 \cdot \pi \cdot G}} \cdot t \cong \tilde{\phi}_i \cdot \left( \exp(-\tilde{a}_0 \cdot t / \alpha) \approx 1 - \tilde{a}_0 \cdot t / \alpha \right) \qquad (19)$$

If we assume that $\tilde{\phi}_0 \cong \tilde{\phi}_i$, and that the time factors are small, we can state

$$\frac{m}{\sqrt{12 \cdot \pi \cdot G}} \cong \frac{\tilde{a}_0}{\alpha} \qquad (20)$$

as an order of magnitude estimate for the initial value of our scale factor at the beginning of inflation. So being the case, we move then to obtain a value for the initial evolution of the Hubble parameter via use of conformal time, with an Einstein equation[18]

$$\ddot{\phi} + 2 \cdot a \cdot H \cdot \dot{\phi} + a^2 \cdot m^2 (\phi - \phi^*) = 0 \qquad (21a)$$

where the conformal time roughly goes as[19]

$$\tilde{t} \cong -\frac{1}{a(t) \cdot H} \tag{21b}$$

which may be re written using ordinary time as[20]

$$\ddot{\phi} + 3 \cdot H \cdot \dot{\phi} + m^2 (\phi - \phi^*) = 0 \tag{21c}$$

which would lead to $\in(t) \equiv \lambda_0 / (H(t)^4) \approx 1$, implying a nucleation rate evolution[16] along the lines of

$$H \cong \frac{1}{3} \cdot \left(\frac{\tilde{a}_0}{\alpha}\right)^{-1} + \left(\frac{\tilde{a}_0}{\alpha}\right) \frac{m^2}{3} \cdot \left(1 - \frac{\phi^*}{\phi_0} \cdot \exp(\frac{\tilde{a}_0}{\alpha} \cdot \tilde{t})\right) \tag{22}$$

implying

$$\lambda_0 (t_P + \delta \cdot t) \approx H^4 (t_P + \delta \cdot t) \leq H^4 (t_P) \tag{23}$$

which for small times just past the initial value of $t \equiv t_P + \delta \cdot t$ leads to a nearly stable but decreasing rate of the Hubble parameter right after a nucleation of a universe. This also leads to a phase change in behavior which I claim is motivated by the pre Planck time value of the Hubble parameter being set by ( for times $t \leq t_P$ )

$$H^2 \equiv \frac{8 \cdot \pi}{3} \cdot G \cdot V(\phi) \rightarrow \frac{8 \cdot \pi}{3} \cdot V(\phi) \tag{24}$$

with the potential given by Eq. 7 prior to the topological charge argument leading to a potential proportional to $(\phi - \phi^*)^2$, which blends into Guths chaotic inflationary model for times $t_P + \delta \cdot t$.

## VI. STRING THEORY AND THE BEHAVIOR OF OUR SCALAR FIELD $\phi$

I can refer to a basic relationship between our scalar field $\phi$ and the strength of all forces gravitational and gauge alike via a relationship given by Veneziano[21]

$$l_P^2 / \lambda_S^2 \approx \alpha_{GAUGE} \approx e^{\phi} \qquad (25)$$

where the weak coupling region would correspond to where $\phi << -1$ and $\lambda_S$ is a so-called quanta of length, and $l_P \equiv c \cdot t_P \sim 10^{-33} cm$. As Veneziano implies in his Fig.2,[21] a so-called scalar dilaton field with these constraints would have behavior seen by the right-hand side of his Fig. 2, with the $V(\phi) \to \varepsilon^+$ but would have no guaranteed false minimum $\phi \to \phi_F < \phi_T$ and no $V(\phi_T) < V(\phi_F)$. The typical string models assume that we have a present equilibrium position in line with strong coupling corresponding to $V(\phi) \to V(\phi_T) \approx \varepsilon^+$ but no model corresponding to potential barrier penetration from a false vacuum state to a true vacuum in line with Colemans presentation[12]. However, FRW cosmology[22] will in the end imply the following

$$t_P \sim 10^{-42} \sec onds \Rightarrow size \; of \; universe \approx 10^{-2} cm \qquad (26)$$

which is still huge for an initial starting point, whereas we manage to in our S-S' distance model to imply a far smaller but still non-zero radii for the initial 'universe' in our model.

## VII. CONCLUSION

This article proposes the case that the false vacuum hypothesis[12] provides a necessary condition for considering transport between adjacent (but varying in magnitude) local minimum values for the generalized potential that are split into the three parts. For the first potential, the difference in the relative energy levels of the local minimum leads, by the Bogomil'nyi inequality,[5,12] to conditions of a nucleating S-S' pair. The second potential corresponds to a decaying scalar field but initial potential hill corresponding to the topological charge vanishing in the onset of inflation, whereas the third potential is the blend into traditional chaotic potentials as outlined by Guth[3] and colleagues as a model for how chaotic inflation corresponds to approaching a global minimum as a ground state of inflationary cosmology that is observable. I assume that the typical Heisenberg uncertainty principle, with respect to uncertainty in time and energy values leads to an initial pop up of a scalar potential to be in the neighborhood of the minimum of Figure 1b, whereas afterwards topological charge of a nucleating potential pair vanishes after this a prioriti collapse, leading to a reduction of scalar potential values, and leading to the behavior as designated by the second and third potentials given in Eqs. 10a,b to 13 above.

I present an argument for a newly nucleated universe to have a finite (but quite small) diameter and reconcile the chaotic inflationary model of Guth[3] with a new fate of the false vacuum paradigm for nucleation at the initial stages of the big bang. In addition, setting the magnitude of Plancks time $t_P = \left(\frac{\hbar \cdot G}{c^5}\right) \to 1$ is part and parcel of how to set up an evolving nucleation rate per Hubble-volume per Hubble-time. I find a way to scale the behavior of the nucleation rate with the behavior of a time by varying initial Hubble parameter for times $t_P + \delta \cdot t$, which more precisely represents the initial behavior of inflationary cosmology. I am assuming for now that the initial radius of a nucleating universe is largely answered by[21] $l_P^2 / \lambda_S^2 \approx \alpha_{GAUGE} \approx e^\phi$, with the scalar dilaton field give by the arguments presented, with a peak value of $\phi \approx 2 \cdot \pi$.

Next, we need research into how the behavior of a domain wall hypothesis, which collapses after Planck time $t_P$, influences the vanishing of topological charge, characterized in this paper as a primary mover in the reduction of our initial nucleation potential in relation to the Guth inflationary potential[3] used in chaotic inflationary cosmology. As is well-known, Frank Wilczek's research[23] into fractional quantum numbers for chiral fermions on so called higher dimensional walls depends upon a counting algorithm pioneered by Schrieffer. The procedure outlined here assumes an initial S-S' configuration with an individual soliton similar to chiral fermions. The application of this detail in relation to this presentation will be more fully delineated in a future paper.

**FIGURE CAPTIONS:**

**FIG 1a.** Evolution from an initial state $\Psi_i[\phi]$ to a final state $\Psi_f[\phi]$ for a double-well potential (inset) in a 1-D model, showing a kink-antikink pair bounding the nucleated bubble of true vacuum. The shading illustrates quantum fluctuations about the initial and final optimum configurations of the field, while $\phi_0(x)$ represents an intermediate field configuration inside the tunnel barrier. The upper right hand side of Fig. 1a shows how the fate of the false vacuum hypothesis gives a difference in energy between false and true potential vacuum values which we tie in with the results of the Bogomil'nyi inequality.

.

**FIG 1b** : Evolution from an initial state $\Psi_i[\phi]$ to a final state $\Psi_f[\phi]$ for a tilted double-well potential in a quasi 1-D cosmological model for inflation, showing a kink-antikink pair bounding the nucleated bubble of true vacuum. .

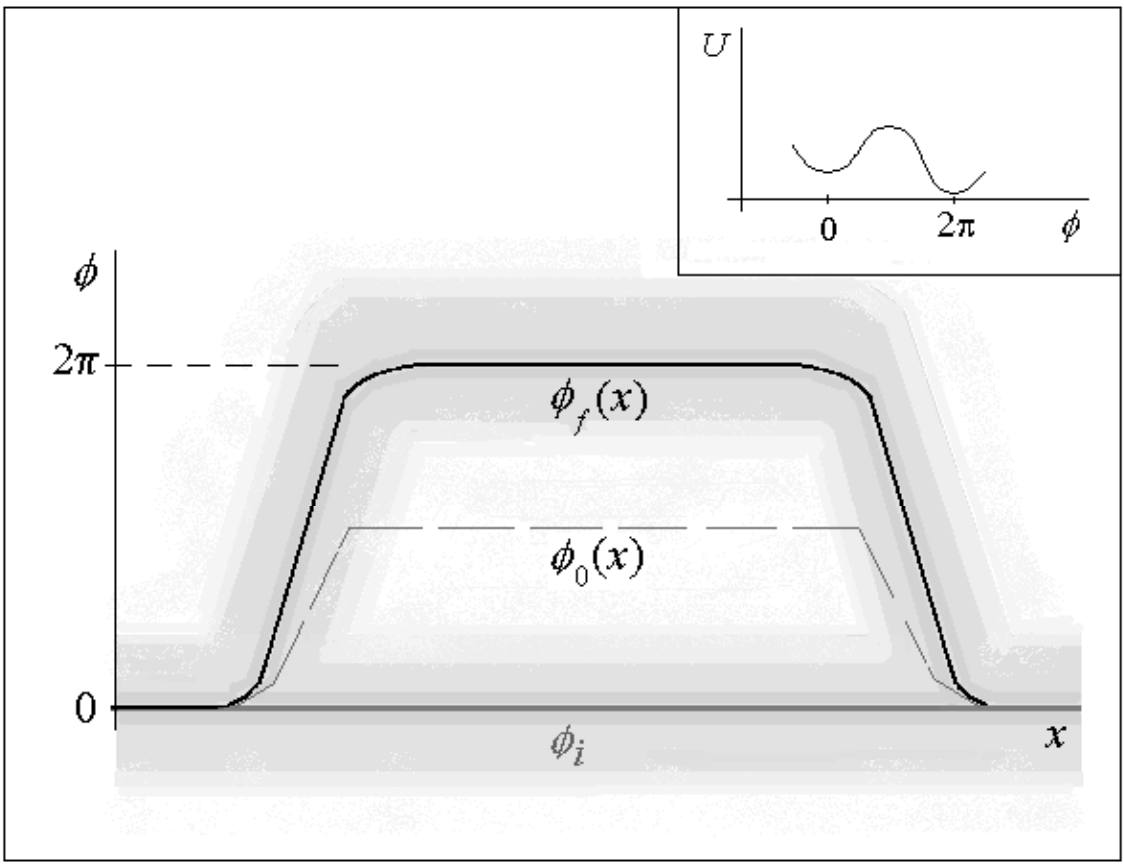

**FIGURE 1A**

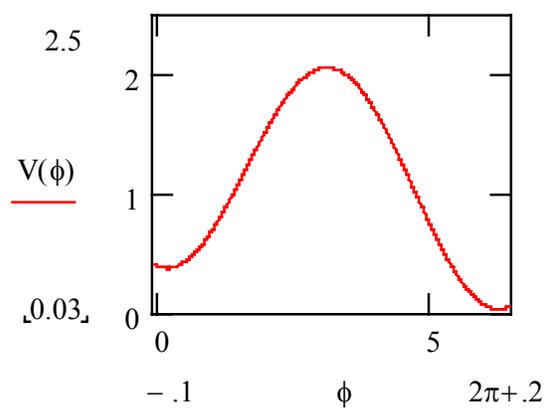

**FIGURE 1B**